\documentclass[a4paper,fleqn]{cas-sc}

\usepackage{natbib}
\usepackage{float} 
\usepackage{stfloats}
\usepackage{booktabs}
\usepackage{bm}

\usepackage{amsmath,amsfonts}
\usepackage{amssymb}
\usepackage{algorithmic}
\usepackage{algorithm}
\usepackage{array}
\usepackage[caption=false,font=normalsize,labelfont=sf,textfont=sf]{subfig}
\usepackage{textcomp}
\usepackage{url}
\usepackage{verbatim}
\usepackage{graphicx}
\usepackage{dsfont}
\usepackage{footnote}
\makesavenoteenv{tabular}
\makesavenoteenv{table}
\usepackage{multirow}
\usepackage{multicol}
\newcolumntype{d}[1]{D..{#1}}
\usepackage{tcolorbox}
\usepackage{flushend}

\usepackage{enumitem}
\def\tsc#1{\csdef{#1}{\textsc{\lowercase{#1}}\xspace}}
\tsc{WGM}
\tsc{QE}
\tsc{EP}
\tsc{PMS}
\tsc{BEC}
\tsc{DE}

\begin{document}
\let\WriteBookmarks\relax
\def\floatpagepagefraction{1}
\def\textpagefraction{.001}

\shorttitle{}

\shortauthors{Miao et~al.}

\title [mode = title]{Adapting General Disentanglement-Based Speaker Anonymization for Enhanced Emotion Preservation}    

\author[1]{Xiaoxiao Miao}[orcid=0000-0002-6645-6524]
\ead{xiaoxiao.miao@singaporetech.edu.sg}
\cormark[1]
\author[2]{Yuxiang Zhang}[orcid=0000-0003-4624-5663]
\ead{zhangyuxiang@hccl.ioa.ac.cn}
\author[3]{Xin Wang}[orcid=0000-0001-8246-0606]
\ead{wangxin@nii.ac.jp}
\author[4]{Natalia Tomashenko}[orcid=0000-0002-7125-2382]
\ead{natalia.tomashenko@inria.fr}
\author[1]{Donny Cheng Lock Soh}[orcid=0000-0002-5505-7501]
\ead{donny.soh@singaporetech.edu.sg}
\author[1]{Ian Mcloughlin}[orcid=0000-0001-7111-2008]
\ead{ian.mcloughlin@singaporetech.edu.sg}

\address[1]{Singapore Institute of Technology, Singapore, 567739}
\address[2]{Key Laboratory of Speech Acoustics and Content Understanding, Institute of Acoustics, Chinese Academy of Sciences, Beijing 100190, China}
\address[3]{National Institute of Informatics, Chiyoda-ku, Tokyo 101-8340, Japan}
\address[4]{Inria, Centre Inria de l'Université de Lorraine, France}

\cortext[cor1]{Corresponding author}
\cortext[cor2]{Xiaoxiao Miao and Yuxiang Zhang contributed equally to this work.}
\cortext[cor3]{This study is supported by JST, PRESTO Grant JPMJPR23P9, Japan, Ministry of Education, Singapore, under its Academic Research Tier 1 (R-R13-A405-0005) and its SIT's Ignition grant (STEM) (R-IE3-A405-0005) andin part by National Research Foundation Singapore for (a) AI Singapore Programme (award AISG2-GC-2022-004) (b) with Infocomm Media Development Authority (Digital Trust Centre award DTC-RGC-07).}

\begin{abstract}
A general disentanglement-based speaker anonymization system typically separates speech into content, speaker, and prosody features using individual encoders. This paper explores how to adapt such a system when a new speech attribute, for example, emotion, needs to be preserved to a greater extent.
Two strategies for this are examined.
First, we show that integrating emotion embeddings from a pre-trained emotion encoder can help preserve emotional cues, even though this approach slightly compromises privacy protection.
Alternatively, we propose an emotion compensation strategy as a post-processing step applied to anonymized speaker embeddings. This conceals the original speaker's identity and reintroduces the emotional traits lost during speaker embedding anonymization.
Specifically, we model the emotion attribute using support vector machines to learn separate boundaries for each emotion. During inference, the original speaker embedding is processed in two ways: one, by an emotion indicator to predict emotion and select the emotion-matched SVM accurately; and two, by a speaker anonymizer to conceal speaker characteristics. The anonymized speaker embedding is then modified along the corresponding SVM boundary towards an enhanced emotional direction to save the emotional cues. 
The proposed strategies are also expected to be useful for adapting a general disentanglement-based speaker anonymization system to preserve other target paralinguistic attributes, with potential for a range of downstream tasks\footnote{Emotion compensation code and audio samples are available at \url{https://github.com/xiaoxiaomiao323/emotion-compensation}}.

\end{abstract}


\maketitle

\section{Introduction}\label{intro}

Voice-based human-computer interaction is becoming increasingly prevalent, offering significant convenience in our daily lives. 
In such systems, a standard data flow typically involves speech being uploaded to a cloud service or social media for further analysis.
There are several potential risks associated with directly uploading raw audio to unreliable parties or social media. 
Various privacy-related traits about an individual can be derived from their voice, encompassing both paralinguistic and linguistic aspects. 
Regarding paralinguistic features, most concerns centre around speaker identity (such as voice-based biometrics) being used to enable deepfake manipulation, since speaker identity is widely utilized in emerging voice authentication systems \citep{kinnunen2010overview, hansen2015speaker, bai2021speaker}. 
Another area of concern involves deducing additional paralinguistic information from raw speech, leading to the creation of applications such as targeted advertisements based on factors like customer age, gender, and accent \citep{Lawsuit2022}.

Existing privacy-preserving approaches for speech technology, known as speaker anonymization, mainly suppress the speaker’s identity (privacy), leaving the linguistic content and other paralinguistic attributes (such as age, gender, emotion, dialect) unchanged (utility). This facilitates the use of anonymized speech for various downstream tasks. 
The VoicePrivacy Challenge (VPC) launched the voice protection initiative, holding challenges in 2020 \citep{tomashenko2021voiceprivacy}, 2022 \citep{tomashenko2022voiceprivacy, 10603395}, and 2024 \citep{tomashenko2024voiceprivacy}. 
This initiative provided a formal definition of the speaker anonymization task, common datasets, evaluation protocols, and baseline systems, promoting the development of privacy preservation techniques for speech technology.

In the VPC settings, the main objective is to hide speaker identity, while preserving wanted speech attributes for downstream tasks.
For example, the primary downstream task for anonymized speech in all VPCs is automatic speech recognition (ASR) application where it is essential to preserve speech content. In VPC 2020 and 2022, multi-party human conversations \citep{cai2018exploring,miao2021d} were taken as another downstream task, meaning anonymized speech should preserve speaker distinctiveness to the greatest extent. This ensures that all utterances from a given speaker are uttered by the same pseudo-speaker, while utterances from different speakers are uttered by different pseudo-speakers, referred to as \textit{speaker-level} anonymization. In VPC 2024 \citep{tomashenko2024voiceprivacy}, in addition to ASR application, emotion analysis was considered as another downstream task. Preserving emotional traits in anonymized speech became essential, while speaker distinctiveness was unnecessary. In line with target application scenarios, \textit{utterance-level anonymization} \citep{tomashenko2024voiceprivacy} was applied, where different utterances from the same source speaker are independently anonymized, not relying on speaker labels and typically resulting in different pseudo-speakers.

In terms of baseline systems provided by VPCs, there are two main variants, including digital signal processing (DSP)-based and deep neural network (DNN)-based methods.
The DSP-based method is a training-free approach that perceptually modifies the McAdams coefficient \citep{mcadams1984spectral, patino2020speaker} to shift the spectral envelope. 
Similar works have changed speech characteristics from the speech production perspective to hide the original speaker's identity, such as formant \citep{gupta2020design,dubagunta2020adjustable}, voice speed \citep{vpc22tsm}, and other vocal tract and voice source features \citep{tavi2022improving}.
However, many studies have shown that DSP-based methods suffer from content distortion problems and are inefficient against stronger attackers \citep{tomashenko2020introducing, tomashenko2021voiceprivacy, srivastava2020evaluating}.

The techniques of DNN-based speaker anonymization systems (SAS) borrow from the domains of neural voice conversion and speech synthesis, primarily focusing on disentangled representation learning approaches. The primary reason for choosing disentanglement-based methods is their capability to separate linguistic and paralinguistic features using distinct encoders, allowing independent modification of each attribute. Specifically, the primary DNN-based frameworks of VPC 2020 and 2022 initially decompose original speech into content \citep{povey2018semi}, speaker identity \citep{snyder2018x}, and pitch features \citep{kasi2002yet}. 
There is an assumption that speaker identity representation contains the majority of private speaker information. 
Hence, to conceal the original speaker identity information, a mainstream method involves replacing the original speaker embedding with a pseudo-speaker (artificial) embedding. 
To generate a pseudo-speaker embedding, an external original speaker pool with numerous speakers is utilized to identify one or several speaker embeddings, which are then averaged to form an anonymized speaker embedding \citep{fang2019speaker,Srivastava2020DesignCF,srivastava2022privacy}, denoted the selection-based speaker anonymizer.
Finally, the original content and prosody features, along with the anonymized speaker embedding, are fed into a speech generator \citep{wang2019neural, kong2020hifi} to generate anonymized speech.

As DNN-based baselines achieve a good balance between privacy and utility \citep{fang2019speaker,Srivastava2020DesignCF,srivastava2022privacy}, many studies following the VPC protocol have proposed improvements from various perspectives. These include: (i) enhancing disentanglement to mitigate the leakage of speaker privacy information from original content and prosody features \citep{shamsabadi2022differentially, mawalim2022speaker, champion2022disentangled, champion2023anonymizing, yao2024musa}, (ii) improving the speaker anonymizer to generate natural and diverse anonymized speaker vectors that protect speaker privacy when facing various threats \citep{meyer2023prosody, miao2023language, yao2024distinctive}, (iii) adapting anonymization techniques to modify not only speaker identity but also additional privacy-related paralinguistic traits such as age, gender, and accent, thereby enhancing the flexibility of anonymization \citep{zhang2023voicepm, noe2023hiding}, (iv) exploring language-robust speaker anonymization methods using self-supervised learning (SSL) model \citep{hsu2021hubert,baevski2020wav2vec} to enable SAS systems trained primarily on English data to anonymize speech for unseen languages \citep{miao22_odyssey,miao2022analyzing,miao2023language,meyer2024probing}, (v) investigating new strategies under more realistic and complex multi-speaker scenarios rather than single-speaker non-spontaneous scenarios \citep{miao2024benchmark}, and (vi) developing robust and modular speaker anonymization and evaluation frameworks \citep{meyer2023voicepat, zhang2023voicepm, franzreb2023comprehensive}.
In addition to improving the anonymization and evaluation methods, a few studies have focused on creating privacy-friendly training datasets for automatic speaker verification \citep{miao2024synvox2} and speech synthesis \citep{huang2024multi} using speaker-anonymized data. 

Besides the baseline systems introduced above, VPC 2024 provides additional DNN-based SASs. These include systems using phonetic transcriptions and a generative adversarial network \citep{meyer2023prosody}, neural audio codec language modeling \citep{panariello_speaker_2023}, and vector quantization \citep{champion2023anonymizing}.
Unfortunately, the VPC 2024 results for all these SASs (referred to as general\footnote{'general' refers that during anonymization, the focus is typically on speaker identity, content, and prosody, without explicitly considering other paralinguistic attributes} SASs.) revealed a loss of emotion \citep{tomashenko2024voiceprivacy}, as they lack special consideration or mechanisms to preserve emotional status.
There have been recent attempts to hiding speaker identity while preserve emotional states and content, submitted to the VPC 2024. For example, \citep{xinyuan24_spsc} explored voice conversion and cascaded ASR-TTS-based speaker anonymization systems, allowing for flexible adjustment of the trade-off between emotion preservation and anonymization by performing a random admixture of different anonymization systems.
\citep{hua24_spsc} proposed extracting two types of deep representations: one using the WavLM model to extract speech features, followed by a method called GMM-Blender to generate a rich array of anonymous templates from an anonymization pool. The other employs a Wav2vec2-based emotion extraction module to preserve emotional attributes. Both representations were fed into the HiFi-GAN model for high-quality anonymized speech generation.
In another approach, \citep{yao24_spsc} utilized a serial disentanglement strategy to separate linguistic content, speaker identity, and emotional state step by step, enabling effective anonymization while maintaining emotional expressiveness.
\citep{gu24_spsc} disentangles content representation using an ASR-BN extractor and non-content representation using the Global Style Token strategy. The non-content representation is replaced with a pseudo-speaker chosen from an external pool, ensuring the speaker's emotion matches the original utterance.

In this paper, we adhere to the VPC 2024 and primarily used its evaluation protocol to investigate how to improve the preservation of a new speech attribute (namely emotion), given a general disentanglement-based SAS. 
The investigation is done in two ways.
On one hand, we demonstrate that explicitly extracting emotion embeddings from a pre-trained emotion encoder helps preserve emotional cues, but it slightly reduces the privacy protection ability. 
This is because emotion features extracted by a pre-trained emotion encoder inevitably contain speaker characteristics. 
Consequently, information about the speaker's identity can be leaked from the emotional features.
On the other hand, we verify that speaker embeddings, while being designed to capture unique speaker characteristics, often retain additional information \citep{raj2019probing}, including emotional cues \citep{ulgen2024revealing}. Directly replacing or transforming original speaker embeddings without considering emotional attributes will cause a loss of emotional information.

To mitigate this problem, we propose an emotion compensation strategy as a post-processing step to modify the anonymized speaker embeddings. It is designed to reintroduce emotional traits that may have been reduced or lost during the speaker anonymization process. 
Specifically, we assume that in the latent space of the (original and anonymized) speaker embeddings, there is a direction for manipulating each basic type of emotion such as \emph{happy}. If we move the speaker embedding along the direction, the reconstructed speech using the speaker embedding will gradually change from a happy voice to an unhappy one, or vice versa. This assumption is similar to that for face editing \citep{shen2020interpreting} --- by moving the latent face embedding along one specific direction, we morph the face image from being smiling into unsmiling. 
Ideally, for a speaker embedding that becomes `less happy` after anonymization, we compensate for the happiness by manipulating the embedding along the happiness direction in the latent space. 

As an implementation, we use support vector machines (SVMs) to find the latent directions for emotion compensation.
Specifically, we train an SVM for each emotion individually (i.e., happy, angry, neutral, sad) using original speaker embeddings on emotion datasets.
The SVM classifies whether the speaker embedding has the corresponding emotion or not. 
The normal vector is decided by the classification hyperplane service as the direction for emotio compensation. 
During inference, after obtaining the speaker embedding from the original speech, we first anonymize the speaker identity using our recently proposed orthogonal Householder neural network (OHNN)-based speaker anonymizer \citep{miao2023language} to ensure the removal of speaker identity. Meanwhile, we use emotion indicators to predict emotions from the input original speaker embedding. 
We then pick the corresponding emotion SVM and manipulate the anonymized speaker embedding along the normal vector of the SVM hyperplane and compensate for the reduced emotion attribute.

Though the paper focuses on emotional attributes, this strategy could be easily extended to other paralinguistic attributes, such as accent, gender, and so on.
In summary, the proposed strategies are expected to be useful for adapting a general disentanglement-based SAS to preserve another new paralinguistic attribute, making it suitable for different downstream tasks.

\begin{figure}[]
  \centering
\includegraphics[width=6in]{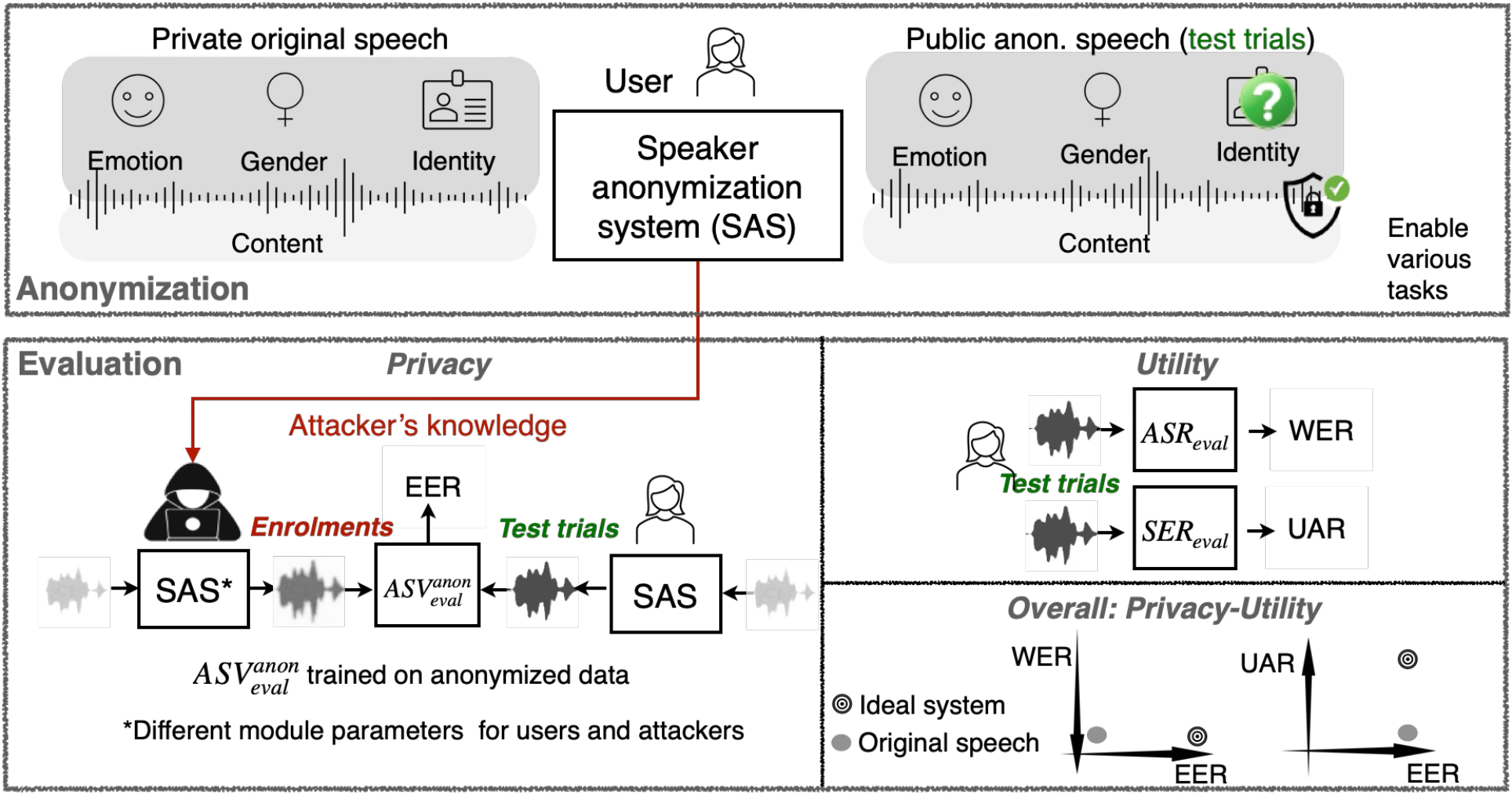}
    \caption{Speaker anonymization task and its evaluation protocol: Users anonymize their original speech to conceal their identities before publication. Meanwhile, attackers use biometric (ASV) technology and their knowledge of the anonymization method to infer the original speaker. 
    The evaluation metrics include equal error rate (EER) of ASV, word error rate (WER) of automatic speech recognition (ASR), and unweighted average recall (UAR) of speech emotion recognition (SER). While the ASV EER measures the goodness of privacy protection, the other two are for utility evaluation. 
    }\label{fig:ssa}
\end{figure}

\section{Related Work}
In this section, we describe the official design for VPC 2024, which serves as the foundation for this study. 
This includes the requirements of the task, threat scenarios, considered downstream tasks, and evaluation metrics. 
Additionally, we provide an overview of the baseline speaker anonymization approaches from VPC 2024, as well as our recently proposed SASs.

\subsection{VoicePrivacy 2024}
\subsubsection{Requirements}
The VPC conceptualizes speaker anonymization as a game between users and attackers, as illustrated in Fig. \ref{fig:ssa}.
Users anonymize their speech (referred to as \emph{test trial} utterances) using an SAS to maximize privacy protection by concealing their identities, while still maintaining the utility required for downstream tasks such as ASR and emotion analysis. 
Meanwhile, attackers with prior knowledge attempt to infer the original speaker identities from the anonymized speech published by users.
According to the VPC 2024 evaluation plan \citep{tomashenko2024voiceprivacy}, an SAS should:
\begin{itemize}
\item take  original speech as input and generates anonymized speech;
\item conceal the speaker's identity;
\item preserve the linguistic and emotional traits.
\end{itemize}
The anonymized speech sounds as if it was spoken by another speaker, called a \emph{pseudo-speaker}. 
This \emph{pseudo-speaker} is generated independently for each utterance and may be an artificial voice not associated with any real person.

\subsubsection{Privacy Evaluation}
The bottom left side of Fig. \ref{fig:ssa} illustrates the privacy evaluation processing, where an attacker has access to the original speech of users and knows the structure of the SAS used by the user, albeit without the exact parameters. 
The actions of an attacker are as follows:
\begin{itemize}
\item employs a similar SAS to anonymize the original speech of users, referred to as \emph{enrollment} utterances;
\item pairs anonymized \emph{enrollment} utterances with anonymized \emph{test trial} utterances to guess the speaker identity from the anonymized  \emph{test trial} utterances;
\item establishes an attacker model to measure the similarity for each pair. VPC 2024 considers a strong attacker system, denoted as {$ASV_\text{eval}^\text{anon}$}. The structure is ECAPA-TDNN \citep{desplanques2020ecapa}, trained using extra anonymized data (i.e. \textit{LibriSpeech-train-clean-360} \citep{panayotov2015librispeech}) to reduce the mismatch between original and anonymized speech for more accurate verification/attacking.
\end{itemize}
The attacker uses the {$ASV_\text{eval}^\text{anon}$} model to compute the similarity (i.e., the ASV score) for each enrollment-test trial pair, determining same-speaker and different-speaker by thresholding. 
If the utterances originate from the same speaker, after anonymization, the goal of the users is to achieve a low ASV score to mislead attackers. However, leakage of speaker identity from \emph{test trial} would result in a higher-than-threshold ASV score. Denoting by $P_\text{fa}(\theta)$ and $P_\text{miss}(\theta)$ the false alarm and miss rates at threshold~$\theta$, the EER metric corresponds to the threshold $\theta_\text{EER}$ at which the two error rates are equal, i.e., $\text{EER}=P_\text{fa}(\theta_\text{EER})=P_\text{miss}(\theta_\text{EER})$. A higher EER indicates greater privacy, while a lower EER indicates worse privacy protection. In the rest of the paper, we exclusively use EER to refer to the EER of the attacker's ASV system.

\subsubsection{Utility Evaluation}

The utility evaluation depends on the downstream tasks and VPC 2024 considers ASR and emotion analysis. 
The evaluation of speech content and emotion preservation of anonymized \emph{test trial} speech is straightforward shown in the bottom right side of Fig. \ref{fig:ssa}.

The speech content preservation ability in anonymized speech is assessed by the word error rate (WER) computed using an ASR evaluation model\footnote{\url{https://huggingface.co/speechbrain/asr-wav2vec2-librispeech}} denoted as $ASR_\text{eval}$. It was fine-tuned on \textit{LibriSpeech-train-960} from \textit{wav2vec2-large-960h-lv60-self}\footnote{\url{https://huggingface.co/facebook/wav2vec2-large-960h-lv60-self}}. A lower WER, similar to that of the original speech, indicates good speech content preservation ability.
The emotion preservation ability in anonymized speech is assessed by unweighted average recall (UAR) produced by a pre-trained speech emotion recognizer (SER), 
which is a wav2vec2-based system. A higher UAR, similar to that of the original speech, indicates better emotional content preservation ability. {In the rest of the paper, we exclusively use UAR to refer to the UAR of the SER system.

\subsubsection{Overall Privacy-Utility Evaluation}

In summary, a good SAS should achieve a high EER (close to 50\%), a WER as low as that of the original speech, and a UAR as high as that of the original speech, as plotted in the bottom right corner of Fig. \ref{fig:ssa}.

\begin{table}[t]
\centering
\caption{Summary of investigated SAS systems. \textbf{B1-B6} are VPC 2024 baselines, \textbf{Se} and \textbf{OH} are latest language-independent systems, and \textbf{P1-P3} are proposed emotion-enhanced SASs built upon \textbf{OH}.}
\label{tab:notations}
\resizebox{1\linewidth}{!}{
\begin{tabular}{clccccc}
\toprule
 & Notation &  \makecell[c] {Prosody \\ extractor} & \makecell[c] {Content \\ encoder} &  \makecell[c]{Speaker  \\encoder}  & \makecell[c]{Syn.\\ model}   & \makecell[c] {Speaker \\ anon.} \\\midrule

\multirow{10}{*}{\rotatebox{90}{Disentangle}}&  \textbf{B1}  &    {YAAPT}    & {TDNN-F}  & {TDNN}   &{NSF+HiFi-GAN}  & {Select}   \\  
  &  \textbf{B3}       &  {Phone aligner, Praat }  & {E2E ASR}  & {GST}   &{FastSpeech2 + HiFi-GAN}  & {GAN}   \\ 
  &  \textbf{B5}    &  YAAPT     & wav2vec2 +TDNN-F+ VQ &  {ECAPA}   &{HiFi-GAN}  & {Select}   \\ 
  &  \textbf{B6}   &      YAAPT  &  {TDNN-F+ VQ} &  {ECAPA}  &{HiFi-GAN}  & {Select}   \\ \cmidrule(lr){2-7}

               &     \textbf{Se}   &YAAPT    & {SSL}  & {ECAPA}   & {HiFi-GAN}  & {Select}   \\ 

                &       \textbf{OH}      &YAAPT        & {SSL}  & {ECAPA}   & {HiFi-GAN}  &   \makecell[c]{OHNN}    \\
                &  \textbf{P1}  & \multicolumn{5}{l}{    \cellcolor{gray!7}{\textbf{OH}  + Emotion Encoder}} \\
                & \textbf{P2} & \multicolumn{5}{l}{  \cellcolor{gray!7}{\textbf{OH}  + Emotion Compensation}} \\
                & \textbf{P3} &  \multicolumn{5}{l}{ \cellcolor{gray!7}{\textbf{OH} + Emotion Encoder + Emotion Compensation}} \\
  \midrule

Codec  &  \textbf{B4}       &    \multicolumn{5}{c}{HuBERT Base quantized semantic encoder + EnCodec encoder and decoder, Selection-based speaker anonymizer }   \\ 
\midrule

DSP & \textbf{B2} &  \multicolumn{5}{c}{McAdams coefficients-based} \\
\bottomrule
\end{tabular}
}
\end{table}

\begin{figure}[!t]
 \centering
 \subfloat[A General disentanglement-based SAS]{\includegraphics[width=0.30\textwidth]{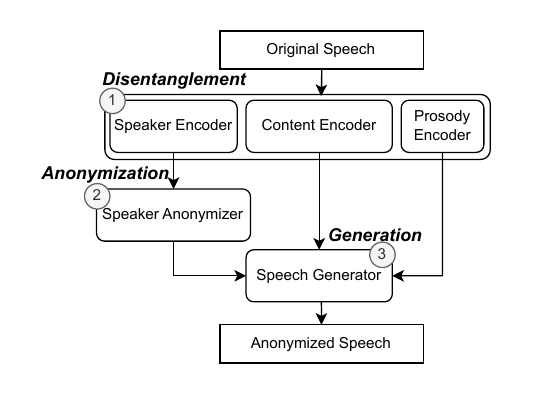}\label{fig:gen-sas}}
 \quad
 \subfloat[\textbf{Se}: SSL-based SAS using selection-based speaker anonymizer]{\includegraphics[width=0.3\textwidth]{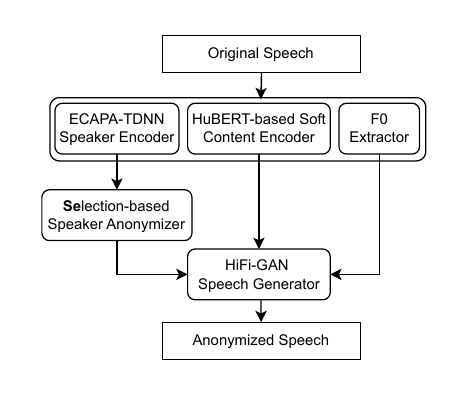}\label{fig:ssl-sas}}
 \quad
 \subfloat[\textbf{OH}: SSL-based SAS using OHNN-based speaker anonymizer]{\includegraphics[width=0.30\textwidth]{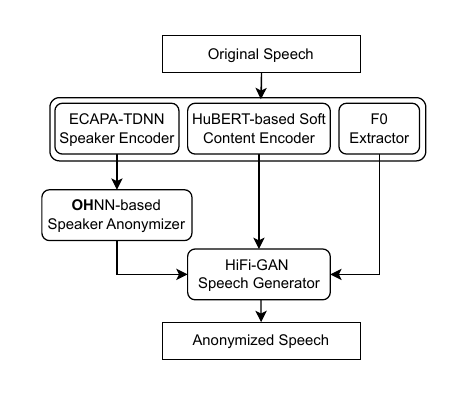}
 \label{fig:ssl-sas-ohnn}} \\ 
 \subfloat[\textbf{P1}: \textbf{OH} with a pre-trained emotion encoder]{\includegraphics[width=0.30\textwidth]{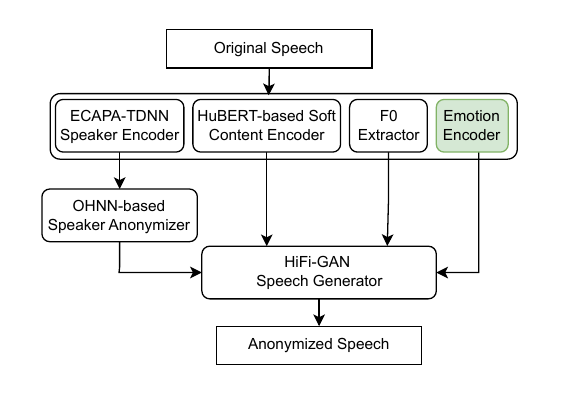}\label{fig:oh-em}}
 \quad
 \subfloat[\textbf{P2}: \textbf{OH} with the emotion compensation]{\includegraphics[width=0.30\textwidth]{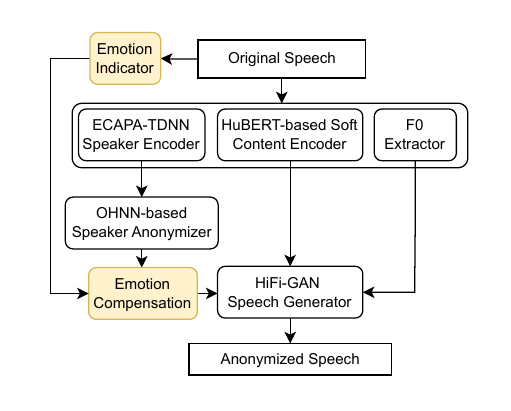}\label{fig:oh-ec}}\quad
 \subfloat[\textbf{P3}: \textbf{OH} with emotion encoder and emotion compensation]{\includegraphics[width=0.30\textwidth]{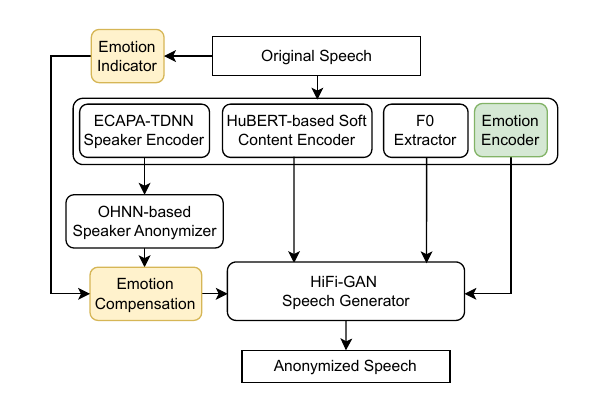}\label{fig:oh-em-ec}}
 
\caption{Disentanglement-based speaker anonymization systems and the proposed emotion-enhanced systems built upon \textbf{OH}. Note that the SSL-based \textbf{Se} and other disentanglement-based systems have a similar structure to \textbf{OH} but use different sub-modules (see summary in Table~\ref{tab:notations}).}
 \label{fig:basic}
\end{figure}

\subsubsection{Baseline Systems}
Table \ref{tab:notations} lists the configurations of various SASs that this paper compares, including six baseline systems introduced by VPC 2024 (\textbf{B1-B6}), more recent proposed language-robust SASs (\textbf{Se, OH}). The proposed emotion-enhanced SASs (\textbf{P1-P3}) are also listed, and they are explained in Section~\ref{sec:exp}.
All the SASs can be categorized into DSP- and DNN-based systems. More specifically, the DNN-based systems can be divided into disentanglement-based and neural-codec-based approaches:
\begin{itemize}
\item DSP-based \textbf{B2}: uses the McAdams coefficient \citep{mcadams1984spectral, patino2020speaker} to shift the pole positions derived from linear predictive coding analysis of speech signals to change the speaker identity.
\item Disentanglement-based \textbf{B1, B3, B5, B6}: these standard disentanglement-based methods consist of three steps as shown in Fig. \ref{fig:gen-sas}:
(i) disentanglement of original speech: includes a speaker encoder, content encoder, and prosody encoder;
(ii) anonymization of original features: typically modifies speaker vectors using a speaker anonymizer;
(iii) generation of anonymized speech: involves a speech synthesis model. 
In particular:
\begin{itemize}
\item \textbf{B1}, proposed in \citep{fang2019speaker}, using the YAAPT algorithm \citep{kasi2002yet}, x-vector \citep{snyder2018x}, and TDNN-F-based acoustic model \citep{povey2018semi} to extract F0, speaker embeddings, and content features, respectively. Using a selection-based speaker anonymizer to hide original speaker identity and neural source-filter (NSF) + HiFi-GAN based speech generator.
\item \textbf{B3}, proposed in \citep{meyer2022anonymizing}, is an ASR+TTS based system. It uses a CNN-LSTM-based phone aligner and the Praat algorithm \citep{boersma2001praat} to extract prosody features, an adapted global style tokens (GST) model \citep{wang2018style} to extract speaker embeddings, and an end-to-end (E2E) ASR model \citep{watanabe2017hybrid, peng2022branchformer} to extract content features. For speaker privacy anonymization, the system uses a generative adversarial network (GAN) to create artificial speaker embeddings. It also modifies prosody by value-wise multiplication of F0 to mitigate speaker identity leakage from F0. For the speech generator, it employs Fastspeech2 \citep{chien2021investigating} and HiFi-GAN.

\item \textbf{B5, B6}, proposed in \citep{champion2023anonymizing}, share similar components. They use the YAAPT algorithm \citep{kasi2002yet} to extract prosody features, ECAPA-TDNN \citep{desplanques2020ecapa} to extract embeddings, randomly select a target speaker from an external speaker pool as a pseudo-speaker to conceal the original speaker identity, and use HiFi-GAN to generate anonymized speech. The only difference lies in their content feature extractors: \textbf{B5} uses wav2vec2 and TDNN-F, while \textbf{B6} uses only TDNN-F. Both approaches incorporate vector quantization (VQ) on content features to reduce speaker characteristics encoded in content features, thereby enhancing the disentanglement ability.
 \end{itemize}

\item Codec-based \textbf{B4}, proposed in \citep{panariello_speaker_2023}, leverages the capabilities of a neural audio codec (NAC). It is an encoder-decoder neural network where original speech is represented by acoustic tokens encoding speaker characteristics and semantic tokens encoding the spoken content of the utterance. Similar to \textbf{B5} and \textbf{B6}, it selects a pseudo-speaker and extracts the acoustic tokens of the pseudo-speaker to replace those of the original speech.
\end{itemize}

In addition to the VPC 2024 baseline systems, two recently proposed disentanglement-based language-robust SASs are selected as additional baselines.
\begin{itemize}
\item SSL-based SAS using a selection-based speaker anonymizer, denoted as \textbf{Se} and plotted in Fig \ref{fig:ssl-sas}: Proposed in \citep{miao22_odyssey}, the novelty compared with \textbf{B1, B3, B5, B6} is that it replaces the language-specific (English) ASR acoustic model with a language-robust HuBERT-soft model \citep{van2021comparison} to enable anonymization in unseen languages.
\item SSL-based SAS using orthogonal Householder neural network (OHNN), denoted as \textbf{OH} and plotted in Fig. \ref{fig:ssl-sas-ohnn}: Proposed in \citep{miao2023language}. The only difference compared to \textbf{Se} is the speaker anonymizer, which is based on OHNN, rotating the original speaker vectors into anonymized speaker vectors while aiming at maintaining the distribution over the original speaker vector space to preserve the naturalness of the original speech.
\end{itemize}

It is worth noting that all the SASs introduced above are designed based on VPC 2020 and 2022, where they focus on ASR and multi-party communication as downstream tasks without considering emotion content maintenance capability. Therefore, no explicit emotion-related mechanisms are applied.
Among these systems, \textbf{OH} achieves the best trade-off between privacy and utility under the VPC 2022 evaluation protocol \citep{miao2023language}. 
Hence, we take \textbf{OH} as the starting point in the following section to improve emotion content preservation.

\begin{figure}
  \centering
\includegraphics[width=3.5in]{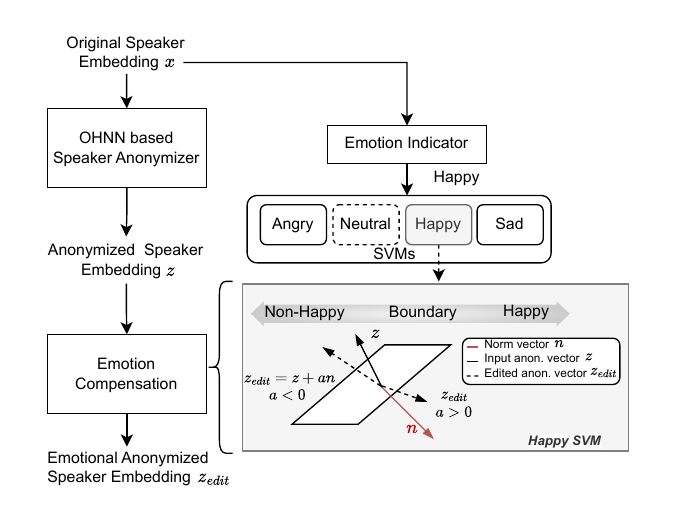}
\caption{The emotion compensation procedure. First, the original speaker embedding \(\mathbf{x}\) is classified by an emotion indicator to select the appropriate SVM (e.g., happy). The embedding \(\mathbf{x}\) is then anonymized as \(\mathbf{z}\). Finally, emotion compensation is performed by $\mathbf{z} + \alpha \mathbf{n}$, where $\mathbf{n}$ is the normal vector corresponding to the hyperplane of the `happy' SVM. 
}
\label{fig:ec}
\end{figure}

\section{Emotion-enhanced OHNN-based Speaker Anonymization Systems}
\label{sec:proposed}

This section introduces strategies for enhancing emotion preservation. 
The first strategy is to add a pre-trained emotion encoder to the disentanglement-based \textbf{OH} SAS (see Fig \ref{fig:oh-em}). By extracting an emotion feature embedding from the input waveform and using it for anonymized speech generation, this strategy is expected to preserve the emotion to some extent.

The second strategy is emotional compensation (see Fig. \ref{fig:oh-ec}).
When disentangling the speech into content, prosody, and speaker features, the emotion is likely to be distributed in the `disentangled' features, e.g., in speaker embeddings~\citep{ulgen2024revealing}. While the emotional traits in the content and prosody features are not changed during the anonymization process, those encoded in the speaker embedding may be lost when the speaker embedding is transformed during anonymization.
The strategy hence is to apply post-processing to the anonymized speaker embedding and recover the lost emotion traits.
Let $\mathbf{x}\in\mathcal{Z}\subseteq\mathbb{R}^d$ be an original speaker embedding vector (with $d$ dimensions), and let $\mathbf{z} = \text{OH}(\mathbf{x}) \in\mathbb{R}^d$ be the anonymized embedding vector produced by the OHNN-based speaker anonymizer. We need to find a post-processing function $f_{\mathbf{\theta}}:\mathcal{Z}\rightarrow\mathcal{Z}$, with which the processed anonymized embedding $f_{\mathbf{\theta}}(\mathbf{z})=f_{\mathbf{\theta}}(\text{OH}(\mathbf{x}))$ encodes the same amount of emotion traits as $\mathbf{x}$.

The implementation is based on the technique of latent features editing, which has been applied to face image morphing~\citep{shen2020interpreting}. 
The assumption is that in the latent space $\mathcal{Z}\in\mathbb{R}^d$ shared by $\mathbf{x}$ and $\mathbf{z}$, we can find a hyperplane (or decision boundary) that well separates speaker embeddings with and without a basic type emotion. 
Let the normal vector $\mathbf{n}\in\mathbb{R}^d$ define a hyperplane of, for example, the happy emotion. The assumption is equivalent to saying that $\mathbf{x}$ from a waveform with happy emotion satisfies $\mathbf{n}^\top\mathbf{x} > 0$, while $\mathbf{x}$ with other emotion has $\mathbf{n}^\top\mathbf{x} < 0$. This is illustrated in Fig.~\ref{fig:ec}.
The value of $\mathbf{n}^\top\mathbf{x}$ is a directional distance from $\mathbf{x}$ to the hyperplane. A larger (and positive) value indicates a stronger emotion.

Hence, to compensate for the emotion associated with a normal vector $\mathbf{n}$, we define a simple post-function $f_{\mathbf{\theta}}$ as 
\begin{equation}
\label{eq:compensation}
    f_{\mathbf{\theta}}(\mathbf{x})\coloneq\mathbf{x} + \alpha\mathbf{n},
\end{equation}
where $\mathbf{\theta} = (\alpha, \mathbf{n})$ is the parameter set that consists of the normal vector of the hyperplane and a hyperparameter $\alpha$ that controls the degree of compensation. Note that $\mathbf{n}\top f_{\mathbf{\theta}}(\mathbf{x}) = \mathbf{n}\top\mathbf{x} + \alpha \mathbf{n}\top\mathbf{n} > \mathbf{n}\top\mathbf{x}$, given $\mathbf{n}\neq \mathbf{0}$ and $\alpha>0$. In other words, the output of $f_{\mathbf{\theta}}$ is expected to have a stronger emotion when using a large (and positive) $\alpha$. Accordingly, we expect that $f_{\mathbf{\theta}}(\mathbf{z})$ compensates (or enhance) the emotion in the anonymized $\mathbf{z}$.
Note that by setting $\alpha<0$ we nullify the emotion. 

In implementation, 
we assume that each emotion has its hyperplane and use a support vector machine (SVM) to learn it~\citep{shen2020interpreting}.
We first train an SVM for each emotion (happy, neutral, sad, and angry) using original speaker embeddings. For example, when training the happy SVM, use speaker embeddings from happy utterances as positive samples, and embeddings from neutral, sad, and angry utterances as negative samples.
After establishing four emotional separation boundaries (SVMs), we use an emotion indicator (i.e., a well-performing emotion recognition model) to predict the emotion of the original speaker embedding 
$\mathbf{x}$ and choose the corresponding SVM. After that, we compensate the emotion in the anonymized speaker embedding $\mathbf{z}$ and get $\mathbf{z}_\text{edit} = f_{\mathbf{\theta}}(\mathbf{z})$.
Fig. \ref{fig:ec} illustrates the procedure, where the compensation is done for the happy emotion.

Note that the reason for using SVM \citep{hearst1998support} is that SVM is a classic and effective machine learning model, the quest to maximize the inter-class spacing when constructing classification boundaries is an important property of SVM. This `large boundary property’ allows an SVM to obtain a clearer separation boundary in a latent space, making subsequent emotion compensation more effective. In addition, SVM has other advantages, such as efficiently handling high-dimensional features, suitable for data with few samples, and strong generalization ability. An SVM classifier is trained for each emotion for binary classification to determine whether the emotion of the input speech is the corresponding emotion, thus obtaining a separation boundary for each emotion. 

\section{Experiments}
\label{sec:exp}
To evaluate the effectiveness of the proposed emotion encoder integration and compensation strategies, we first compare various general SASs to verify the performance of \textbf{OH}. We then apply the proposed strategies to the \textbf{OH} SAS and verify the improvement in emotion preservation capability and overall privacy-utility performance.

\subsection{Datasets and System Configurations}
Baselines \textbf{B1}-\textbf{B6} are from VPC 2024 official implementations \citep{tomashenko2024voiceprivacy}. 
The \textbf{Se} and \textbf{OH} was built using the following VPC standard datasets \citep{tomashenko2024voiceprivacy}:
an ECAPA-TDNN speaker encoder trained on the \textit{VoxCeleb-2} \citep{chung2018voxceleb2};
a HuBERT-based soft content encoder finetuned from a pre-trained HuBERT Base model\footnote{\url{https://github.com/pytorch/fairseq/tree/main/examples/hubert}} on \textit{LibriTTS-train-clean-100} \citep{zen2019libritts}; and a HiFi-GAN model trained on \textit{LibriTTS-train-clean-100} \citep{zen2019libritts}.
Unlike the selection-based anonymizer, which relies on an additional multi-speaker English dataset (\textit{LibriTTS-train-other-500}) containing data from 1,160 speakers as the external pool, the OHNN-based anonymizers reuse a multi-speaker multi-language dataset (\textit{VoxCeleb-2}), that is used to train the ECAPA-TDNN of the SSL-based SAS \citep{miao22_odyssey} with a random seed of 50 for parameter initialization. We used an additive loss function that combines weighted angular margin softmax and cosine similarity. Full details of the training procedure can be found in \citep{miao2023language}.

Systems \textbf{P1}-\textbf{P3} are based on \textbf{OH}. As listed in Table~\ref{tab:notations} and Fig.~\ref{fig:basic}, \textbf{P1} adds an emotion encoder to extract emotion embedding (i.e., the first strategy explained in section~\ref{sec:proposed}); \textbf{P2} adds an emotion indicator and the proposed emotion compensation functions; \textbf{P3} combines \textbf{P1} and \textbf{P2}. 
The emotion encoder in \textbf{P1} and \textbf{P3} uses a pre-trained universal speech emotion representation model called emotion2vec\footnote{\url{https://huggingface.co/emotion2vec/emotion2vec_base}} to extract a 768-dimensional emotion embedding. This model is pre-trained using self-supervised online distillation \citep{ma2023emotion2vec}. The training dataset includes \textit{LibriSpeech-960} \citep{panayotov2015librispeech} and 262 hours of unlabeled emotion data, including 7h of \textit{IEMOCAP} \citep{busso2008iemocap}, 12.2h of \textit{MELD} \citep{poria2019meld}, 91.9h of \textit{CMU-MOSEI} \citep{zadeh2018multimodal}, 37.3h of \textit{MEAD} \citep{wang2020mead}, and 113.5h of \textit{MSP-Podcast} (V1.8) \citep{martinez2020msp}.

The emotion indicator in \textbf{P2} and \textbf{P3} is emotion2vec+\footnote{\url{https://huggingface.co/emotion2vec/emotion2vec_plus_base}}, which is obtained by fine-tuning the pre-trained emotion2vec model using a large-scale pseudo-labeled dataset to predict 9-class emotions: angry, disgusted, fearful, happy, neutral, other, sad, surprised, and unknown. As \textit{IEMOCAP} used in VPC 2024 includes emotions happy, neutral, sad, and angry, we map disgusted and fearful to sad, surprised to happy, and unknown and other to neutral. The average accuracies predicted by emotion2vec+ on the \textit{IEMOCAP-dev} and \textit{IEMOCAP-test} are 89.66\% and 89.80\%, respectively.

Four emotion SVMs are trained using approximately 11 hours of speech data from the \textit{MSP-IMPROV} dataset \citep{MSP-IMPROV} and about 29 hours from the \textit{ESD} dataset \citep{zhou2021seen}. The accuracy on the development set for the happy, neutral, sad, and angry SVMs is 84.17\%, 83.12\%, 92.60\%, and 92.59\%, respectively.

The same speech generator HiFi-GAN is used in \textbf{Se}, \textbf{OH}, and \textbf{P2}. 
It takes 393-dimensional features as input (192-dimensional speaker embeddings from ECAPA-TDNN, 200-dimensional content features from a HuBERT-based soft content encoder, and 1-dimensional F0 extracted by YAAPT) and is trained on \textit{LibriTTS-train-clean-100}. 
The HiFi-GAN for \textbf{P1} and \textbf{P3} also takes 768-dimensional emotion features as input and is trained in the same way as that for other models~\citep{kong2020hifi}, which means the input dimension for HiFi-GAN is 1161, calculated as $192+200+1+768$. Note that both content and speaker embeddings are separately normalized before feeding into HiFi-GAN using instance normalization~\citep{ulyanov2017inn} for each utterence.

For system performance evaluation, we use the official VPC 2024 development and test sets \citep{tomashenko2024voiceprivacy} include recordings from the \textit{LibriSpeech} and 10-speakers \textit{IEMOCAP} \citep{busso2008iemocap} corpora.
Additionally, emotion2vec and emotion2vec+ utilize the IEMOCAP dataset, which makes the evaluation on IEMOCAP less reliable. To verify the effectiveness of the proposed method on completely unseen data, we selected happy, neutral, sad, and angry utterances from the SAVEE dataset~\citep{savee}, including 60, 120, 60, and 60 utterances, respectively. This dataset serves as the evaluation set for the emotion2vec paper \citep{ma2023emotion2vec} and is excluded from the pre-training stage of emotion2vec.

\subsection{Hyperparameter Choice for Emotion Compensation}
\label{sec:sys-set}

When performing emotion compensation on the speaker embedding, the hyperparameter $\alpha$ determines which side of the edited speaker embedding is located (e.g., happy or not) and the level of emotion.
We follow these steps to experimentally determine $\alpha$: 
(i) use labels from \textit{IEMOCAP} dataset\footnote{In theory, any other emotion datasets with their emotional labels can be used.} as authentic indicators, directly select the correct SVMs;
(ii) feed the utterances to the corresponding emotion SVM to modify the anonymized speaker embedding using different values of $\alpha$;
(iii) the compensated speaker embeddings, along with other features, are fed into HiFi-GAN to generate compensated anonymized speech;
(iv) use the standard UAR computation scripts from VPC 2024 to compute the UAR. 
Finally, $\alpha$ is -35 for sad emotion and 35 for the other emotions\footnote{Section \ref{sec:disscussion} will show that the differences caused by different $\alpha$ values are minimal.}.

As the training data of the OHNN-based speaker anonymizer is trained on \textit{VoxCeleb-2}, which comprises a large portion of neutral speech, the transformation of the OHNN is more likely not to change the emotional state for neutral speech. Hence, we skip the emotion compensation step if the emotion of the original speech predicted by the emotion indicator is neutral (i.e., $\alpha=0$).

\begin{table*}[t]
\caption{
   EER (\%), WER (\%), and  UAR(\%) on the VPC dev and test sets when processed by various baseline speaker anonymization systems.}
\label{tab:baseline-compare}
\setlength{\tabcolsep}{6pt}
  \centering
  \footnotesize
  \begin{tabular}{lcccccccccc}  
\toprule
  & \multicolumn{6}{c}{\textbf{EER,\%}  $\uparrow$}   & \multicolumn{2}{c}{\textbf{WER,\%} $\downarrow$}  & \multicolumn{2}{c}{\textbf{UAR,\%} $\uparrow$}  \\  
  \cmidrule(lr){2-7} \cmidrule(lr){8-9}  \cmidrule(lr){10-11} 
      & \multicolumn{3}{c}{Libri-dev}  & \multicolumn{3}{c}{Libri-test}    & \multicolumn{1}{c}{Libri-dev}  & Libri-test   & \multicolumn{1}{c}{IEMOCAP-dev}    & IEMOCAP-test       \\  
  \cmidrule(lr){2-4}  \cmidrule(lr){5-7}
      & F & M & \cellcolor{gray!7} Avg     & F & M &\cellcolor{gray!7}  Avg   & & & &    \\  
\midrule
      
\textbf{Original}   & \multicolumn{1}{c} {10.51} & 0.93 &  \cellcolor{gray!7} 5.72 & 8.76 & 0.42 & \cellcolor{gray!7}4.59   &\multicolumn{1}{c}{1.80} &  1.85  & \multicolumn{1}{c}{69.08} & 71.06   \\
\midrule

\textbf{B1}    & \multicolumn{1}{c} {10.94} & 7.45 & \cellcolor{gray!7}{9.20} & 7.47 & 4.68 & \cellcolor{gray!7}6.07   & \multicolumn{1}{c}{3.07} & 2.91  & \multicolumn{1}{c}{42.71} & 42.78   \\ 

\textbf{B2}    & \multicolumn{1}{c} {12.91} & 2.05 & \cellcolor{gray!7}{7.48} & 7.48& 1.56 & \cellcolor{gray!7}4.52   & \multicolumn{1}{c}{10.44} & 9.95  & \multicolumn{1}{c}{\textbf{55.61}} & \textbf{53.49}   \\ 

\textbf{B3}   & \multicolumn{1}{c}{ 28.43} & 22.04    & \cellcolor{gray!7} {25.24} &  \multicolumn{1}{c}{ 27.92} &  26.72  & \multicolumn{1}{c}{\cellcolor{gray!7} 27.32 } & 4.29 &  4.35  & 38.09 & 37.57\\ 
\textbf{B4}    & \multicolumn{1}{c} {34.37} & 31.06 &\cellcolor{gray!7} {32.71} & 29.37 & 31.16 & \cellcolor{gray!7}30.26   & \multicolumn{1}{c}{6.15} & 5.90  & \multicolumn{1}{c}{41.97} & 42.78   \\ 
\textbf{B5}    & \multicolumn{1}{c} {35.82} & 32.92 &\cellcolor{gray!7} {34.37} & 33.95 & 34.73 & \cellcolor{gray!7}34.34   & \multicolumn{1}{c}{4.73} & 4.37  & \multicolumn{1}{c}{38.08} & 38.17   \\ 
\textbf{B6}    & \multicolumn{1}{c} {25.14} & 20.96 &\cellcolor{gray!7} {23.05} & 21.15 & 21.14 & \cellcolor{gray!7}21.14   & \multicolumn{1}{c}{9.69} & 9.09  & \multicolumn{1}{c}{36.39} & 36.13   \\ 

\textbf{Se}    & \multicolumn{1}{c} {18.32} & 9.62 &\cellcolor{gray!7} {13.97} & 6.20 & 5.79 & \cellcolor{gray!7}5.99   & \multicolumn{1}{c}{\textbf{2.23}} & \textbf{2.34}  & \multicolumn{1}{c}{46.74} & 46.84   \\ 
\midrule
\textbf{OH}    & \multicolumn{1}{c}{\textbf{44.89}} & \textbf{34.74} &\cellcolor{gray!7}{\textbf{39.92}} & \textbf{39.26} & \textbf{37.64} & \cellcolor{gray!7}{\textbf{38.54}}   & \multicolumn{1}{c}{2.36} & 2.48  & \multicolumn{1}{c}{{47.01}} & {47.37}   \\

\bottomrule 

\end{tabular}
\end{table*}

\subsection{Results Analysis}

\subsubsection{The Performance Comparison of VPC 2024 Baselines and the SSL-based SASs}
First, we examine the performance of \textbf{Se} and \textbf{OH}, compared with baselines \textbf{B1}-\textbf{B6} from VPC 2024.
Table \ref{tab:baseline-compare} lists the EER, WER, and UAR results for all the baseline SASs.
For the EER, \textbf{Se} uses a selection-based speaker anonymizer similar to \textbf{B1}, achieving very poor EER, indicating a high degree of privacy leakage. 
This may be because the average operation in the selection-based speaker anonymizer makes the anonymized speaker vectors similar to each other, which is also supported by other studies~\citep{champion2023anonymizing}. 
When an attacker uses the anonymized data to train $ASV_\text{eval}^\text{anon}$, this model would have a more powerful ability to extract the residual distinguishable speaker information from unchanged content and prosody features in the anonymized speech.
By contrast, OHNN rotates the original speaker vector to obtain diverse pseudo-speakers, making it hard for the attacker to infer the original speaker information from the anonymized speech, achieving the highest EER among all the SASs. These findings have been proven in \citep{miao2023language}.

For the WER,  thanks to the power of the HuBERT-based soft content encoder \citep{van2021comparison} and HiFi-GAN \citep{kong2020hifi}, both \textbf{Se} and \textbf{OH} achieve lower WERs than other SASs.

For the UAR, \textbf{Se} and \textbf{OH} achieve over 46\% UAR for both the \textit{IEMOCAP-test} and \textit{IEMOCAP-dev} datasets, only lower than \textbf{B2}. However, \textbf{B2} has poor ability in privacy preservation (very low EER) and content preservation (high WER). One possible reason for the better UAR performance of \textbf{Se} and \textbf{OH} is that all other DNN-based baselines use language-specific ASR to capture linguistic features, while the HuBERT-based soft content encoder in \textbf{Se} and \textbf{OH} does not rely on any specific language, simultaneously capturing emotional cues \citep{yang21c_interspeech}. 
Furthermore, \textbf{OH} achieves better UAR than \textbf{Se}. A potential reason is that \textbf{Se} replaces the speaker embedding with a pseudo-speaker embedding. If the pseudo-speaker embedding encodes a different type of emotion, the emotion in the input speaker embedding is completely lost after anonymization. By contrast, \textbf{OH} anonymizes the speaker embedding by rotation, preserving the emotional cues to some extent.

Overall, \textbf{OH} performs best among all the evaluation metrics. We thus choose this as the backbone to investigate the strategies introduced in section \ref{sec:proposed}.

 \begin{table}[h!]
\caption{UAR (\%) and class-wise recalls (\%) achieved on data processed by \textbf{OH} and variants of \textbf{OH} with the emotion encoder (EE) and emotion compensation (EC).}\label{tab:ser-results}
  \centering
  \begin{tabular}{l ccc c c c c c c c c c} 
    \toprule 
   &&& \multicolumn{5}{c}{IEMOCAP-dev} &  \multicolumn{5}{c}{IEMOCAP-test} \\   \cmidrule(lr){4-8}  \cmidrule(lr){9-13} 
  &  EE& EC  &  \cellcolor{gray!7}UAR  & Sad & Neutral  & Angry & Happy & \cellcolor{gray!7}UAR  & Sad & Neutral  & Angry & Happy \\
  \midrule
\textbf{Ori.}  && & \cellcolor{gray!7}{69.08} & 63.63 & 65.97 & 79.78 & 66.95 &\cellcolor{gray!7}{71.06} & 72.58 & 71.66 & 72.82 & 67.19 \\
    \midrule

\textbf{OH}   &&& \cellcolor{gray!7}{47.01}	& 6.04	& 54.68	& 70.00	& 57.31 &  \cellcolor{gray!7}{47.37}	& 6.85	& 50.57	& 69.48	& 62.57 \\

 \textbf{P1}    & \checkmark&&  \cellcolor{gray!7}{48.96}	& 6.94	& \textbf{65.22}	& 65.21	& 58.45  &  \cellcolor{gray!7}{50.99}	& 8.65	& \textbf{64.42}	& 66.60	& 64.31 \\
\textbf{P2} &&  \checkmark & \cellcolor{gray!7}{54.51}  &  \textbf{9.50} &  54.68 &  80.88 & \textbf{72.99} & \cellcolor{gray!7}{53.72}  & 11.11 & 50.57  & 79.13 & \textbf{74.06} \\
   \textbf{P3}  & \checkmark& \checkmark  &  \cellcolor{gray!7}\textbf{54.78} & 5.16	& \textbf{65.22}	& \textbf{84.91}	& 63.82 &  \cellcolor{gray!7}\textbf{57.93} & \textbf{11.25}	& \textbf{64.42}	& \textbf{84.51}	& 71.55 \\

  \bottomrule 
  \end{tabular}  
\end{table}
\normalsize

\begin{figure}[!t]
 \centering
 \subfloat[\textbf{Original} speech]{\includegraphics[width=0.3\textwidth]{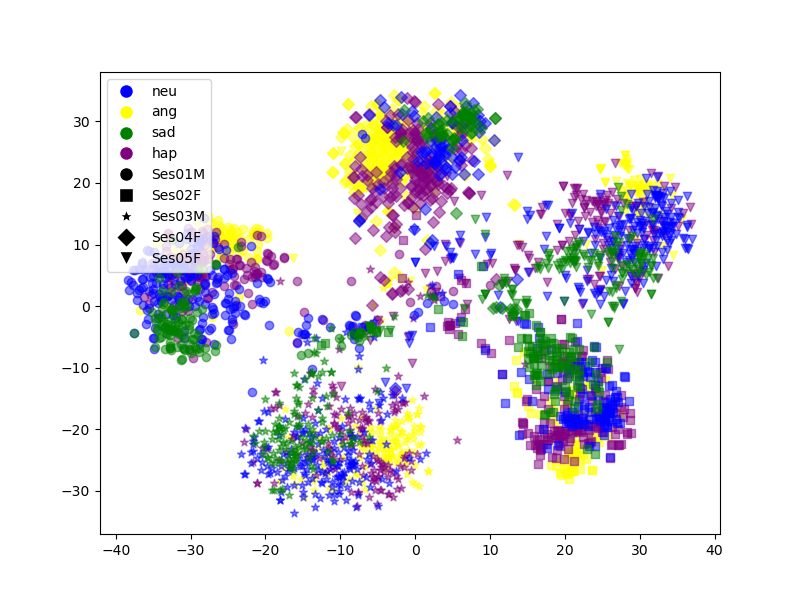}\label{fig:spk-speech-sub1}}
 \hfill
 \subfloat[\textbf{OH} speech]{\includegraphics[width=0.3\textwidth]{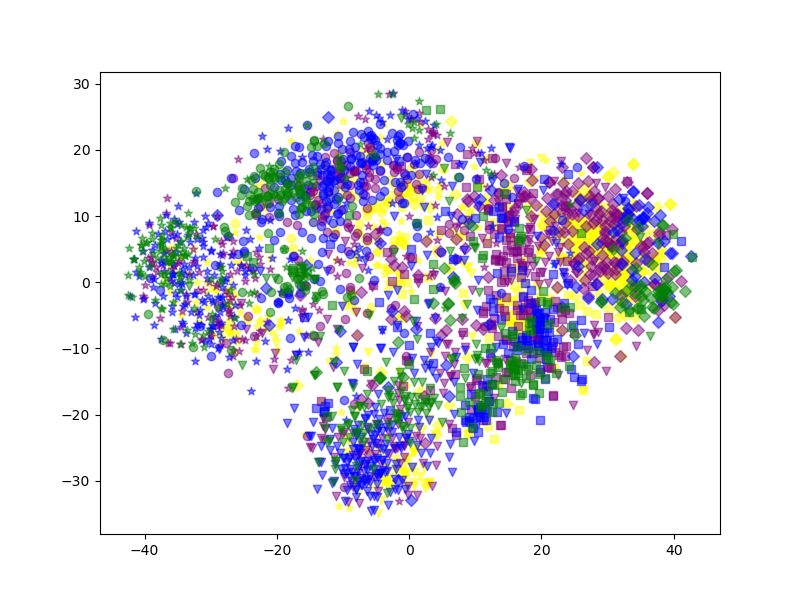}\label{fig:spk-speech-sub2}}
 \hfill
 \subfloat[\textbf{P3}  speech ]{\includegraphics[width=0.3\textwidth]{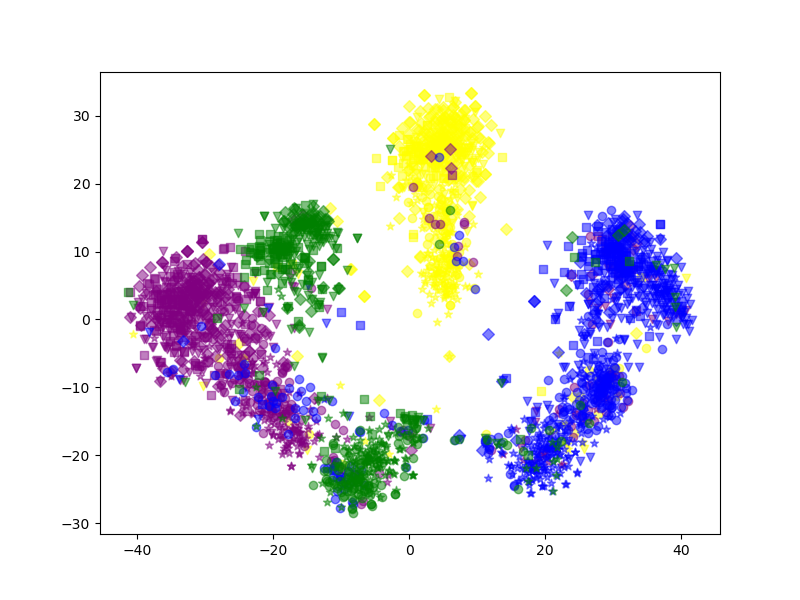}\label{fig:spk-speech-sub3}}
\caption{Speaker embedding visualization from \textbf{original}, \textbf{OH}, and \textbf{P3} speech on \textit{IEMOCAP-test} with 5 speakers and 4 emotions. Colors represent emotions and shapes represent speakers.}
 \label{fig:spk-speech-asv}
\end{figure}

\begin{figure}[!t]
 \centering
 \subfloat[\textbf{Original} speech]{\includegraphics[width=0.3\textwidth]{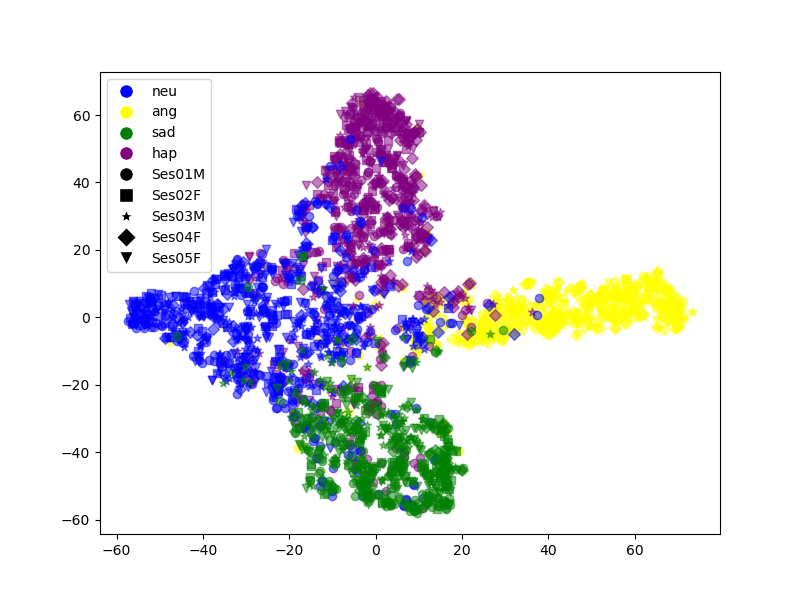}\label{fig:emo-speech-sub1}}
 \hfill
 \subfloat[\textbf{OH} speech]{\includegraphics[width=0.3\textwidth]{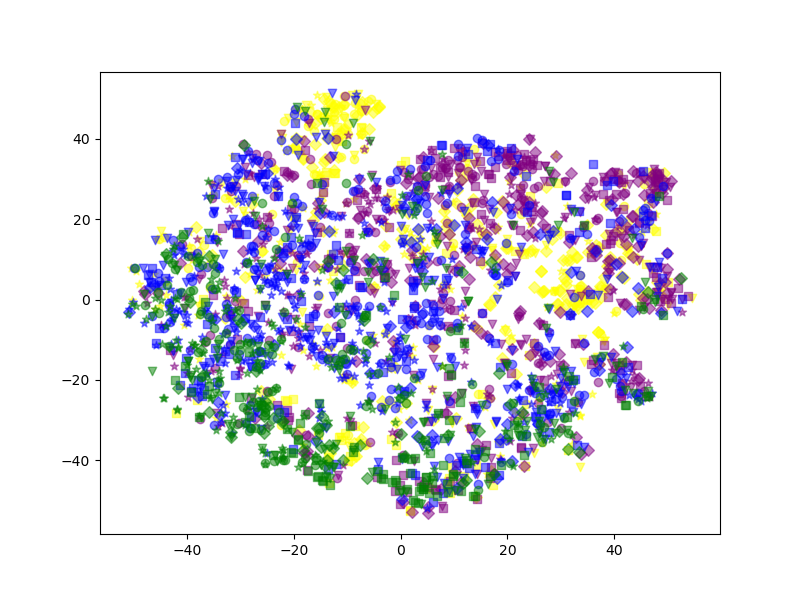}\label{fig:emo-speech-sub2}}
 \hfill
 \subfloat[\textbf{P3} speech ]{\includegraphics[width=0.3\textwidth]{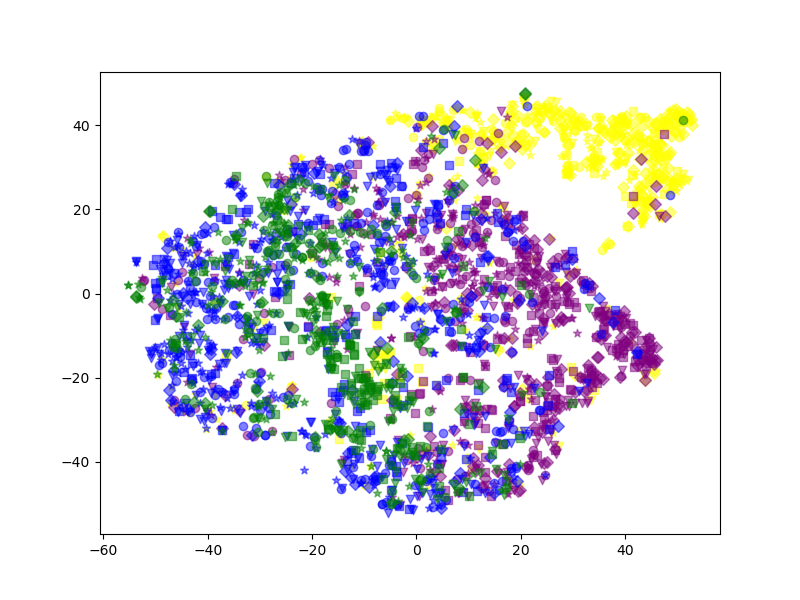}\label{fig:emo-speech-sub3}}
\caption{Emotion embedding visualization from \textbf{original}, \textbf{OH}, and \textbf{P3} speech on \textit{IEMOCAP-test} with 5 speakers and 4 emotions. Colors represent emotions and shapes represent speakers.}
 \label{fig:emo-speech-asv}
\end{figure}

 \begin{table}[h!]
\caption{UAR (\%) and class-wise recalls (\%) achieved on data processed by \textbf{OH} and its variants with the EE and EC using $SER_\text{eval}^\text{emo2vec+}$.}
\label{tab:ser-results2}
\centering
\footnotesize
\begin{tabular}{lccccc|ccccc} 
\toprule 
 &  \multicolumn{5}{c}{IEMOCAP-dev} & \multicolumn{5}{c}{IEMOCAP-test} \\ 
\cmidrule(lr){2-6} \cmidrule(lr){7-11} 
& {UAR} & {Sad} & {Neutral} & {Angry} & {Happy}  & {UAR} & {Sad} & {Neutral} & {Angry} & {Happy} \\ 
\midrule
\textbf{Ori.}  &  \cellcolor{gray!7}{89.62} & 88.07 & 92.55 & 89.10 & 87.83 &  \cellcolor{gray!7}{89.62} & 90.83 & 90.42 & 93.97 & 84.89 \\
\midrule
\textbf{OH}   & \cellcolor{gray!7}{42.53} & \textbf{21.59} &85.22 &14.34 & 28.59 &  \cellcolor{gray!7}{38.53} & 23.92 & 75.81 & 11.38 & \textbf{29.61} \\
\textbf{P1}   &  \cellcolor{gray!7}{43.59} & 15.11 & 84.31 & 15.11 & \textbf{33.21} &  \cellcolor{gray!7}{41.36} & \textbf{24.46} & 80.96 & 24.46 & 29.48 \\
\textbf{P2}    &  \cellcolor{gray!7}{45.67} & 16.67 & 87.97 & 39.20 & 23.48&  \cellcolor{gray!7}{45.89} & 19.60 & 80.84 & \textbf{45.34} & 28.38 \\
\textbf{P3}   & \cellcolor{gray!7}{\textbf{45.74}} & 20.45 & \textbf{90.26} & \textbf{43.42} & 21.90 &  \cellcolor{gray!7}{\textbf{46.21}} & 22.30 & \textbf{85.99} & 41.72 & 24.94 \\
\bottomrule 
\end{tabular}
\end{table}
\normalsize

\subsubsection{The Emotion Preservation Assessment}
Table \ref{tab:ser-results} shows the UAR for different proposed SASs, along with the respective recall rate for each emotion.
By comparing \textbf{OH} and \textbf{P1}, which integrates the emotion encoder into \textbf{OH}, 
we observe that the absolute UAR improvement for \textit{IEMOCAP-dev} and \textit{IEMOCAP-test} is around 2\% (from 47.01\% to 48.96\% and from 47.37\% to 50.99\%, respectively).
This confirms that explicitly using an emotion encoder to extract emotional features is helpful in maintaining more emotional information.
When evaluating \textbf{OH} against \textbf{P2}, which applies post-processing to anonymized speaker embeddings extracted by OHNN with an emotion compensation component, a greater absolute UAR improvement was observed. Specifically, the UAR for \textit{IEMOCAP-dev} and \textit{IEMOCAP-test} increased by around 7\% (from 47.01\% to 54.51\% and from 47.37\% to 53.72\%, respectively).
This proves that speaker embeddings do encode emotional clues, and by modifying speaker embeddings, we can compensate for the lost emotional information.
\textbf{P3} combine both strategies, achieving the best emotion preservation ability among the variants of \textbf{OH}.

To better interpret the UAR improvements, we applied t-distributed stochastic neighbor embedding (t-SNE) \citep{van2008visualizing} to visualize \textit{IEMOCAP-test} speaker embeddings in Fig. \ref{fig:spk-speech-asv} and emotion embeddings in Fig. \ref{fig:emo-speech-asv}. The encoders used here are the same as the corresponding encoders used in \textbf{P3}.
In particular, speaker embeddings from \textbf{original} speech in Fig. \ref{fig:spk-speech-sub1} show five distinct clusters corresponding to the five speakers in the \textit{IEMOCAP-test}, with each cluster containing four emotional sub-clusters (different colors). This demonstrates that the speaker embeddings have both speaker-distinguishing and emotion-distinguishing characteristics.
Speaker embeddings from \textbf{OH} speech in Fig. \ref{fig:spk-speech-sub2} disrupt both speaker and emotion characteristics, while those from \textbf{P3} anonymized speech in Fig. \ref{fig:spk-speech-sub3} has four clear emotion clusters, demonstrating good emotion preservation after applying emotion encoder and emotion compensation.

Similarly, for emotion embeddings visualization, \textbf{original} speech has obvious separation for four speakers in Fig.~\ref{fig:emo-speech-sub1}, which is consistent with the high UAR that the original \textit{IEMOCAP-test} achieved (71.06\%). Unfortunately, it dropped to 47.37\% for \textbf{OH} anonymized speech, which is why Fig. \ref{fig:emo-speech-sub2} does not show any clear separation for emotion. 
However, the emotion embeddings extracted from \textbf{P3} anonymized speech in Fig. \ref{fig:emo-speech-sub3} demonstrate better-separated patterns, particularly for angry (yellow) and happy (purple) emotions, raising the UAR back to 57.93\%. 

Another finding is that the sad emotion (in green color) is much easier to lose and harder to compensate for than the other emotions, as observed from both Table \ref{tab:ser-results} and Fig. \ref{fig:emo-speech-asv}.
If we check the recall on the sad utterances for all the baseline systems in VPC2024 \footnote{\url{https://github.com/Voice-Privacy-Challenge/Voice-Privacy-Challenge-2024/tree/main/results}}, the results suggest that the sad emotion is difficult to preserve, especially for NN-based speaker anonymization methods (B1, B3-B5). Another reason could be that the $SER_\text{eval}$ has a bias against the sad emotion. To investigate further, we applied $SER_\text{eval}^\text{emo2vec+}$ to the IEMOCAP dataset, as shown in Table \ref{tab:ser-results2}. We can see that the recalls on the sad emotion (around 20\%) are higher than those (around 10\%) in Table~\ref{tab:ser-results}. 
This shows that the evaluation model affects the results. 
It also confirms that the sad emotion in the IEMOCAP dataset is hard to preserve in general, even if we use the stronger evaluation model $SER_\text{eval}^\text{emo2vec+}$ that potentially has been exposed to the sad emotion utternaces from the IEMOCAP dataset.

\begin{table*}[t]
\caption{
   EER (\%), WER (\%), and  UAR(\%) on the VPC dev and test sets when processed by various proposed speaker anonymization systems.}
\label{tab:propose-compare}
\setlength{\tabcolsep}{6pt}
  \centering
  \footnotesize
  \begin{tabular}{lcccccccccccc}  
\toprule
  & &&\multicolumn{6}{c}{\textbf{EER,\%}  $\uparrow$}   & \multicolumn{2}{c}{\textbf{WER,\%} $\downarrow$}  & \multicolumn{2}{c}{\textbf{UAR,\%} $\uparrow$}  \\  
  \cmidrule(lr){4-9} \cmidrule(lr){10-11}  \cmidrule(lr){12-13} 
   &  EE& EC  & \multicolumn{3}{c}{Libri-dev}  & \multicolumn{3}{c}{Libri-test}    & \multicolumn{1}{c}{Libri-dev}  & Libri-test   & \multicolumn{1}{c}{IEMOCAP-dev}    & IEMOCAP-test       \\  
  \cmidrule(lr){4-6}  \cmidrule(lr){7-9}
    &&   &  F & M & \cellcolor{gray!7} Avg     & F & M &\cellcolor{gray!7}  Avg   & & & &    \\  
\midrule
      
\textbf{Ori.}  && & \multicolumn{1}{c} {10.51} & 0.93 &  \cellcolor{gray!7} 5.72 & 8.76 & 0.42 & \cellcolor{gray!7}4.59   &\multicolumn{1}{c}{1.80} &  1.85  & \multicolumn{1}{c}{69.08} & 71.06   \\

\midrule

\textbf{OH}   && & \multicolumn{1}{c} {44.89} & 34.74 & \cellcolor{gray!7}{\textbf{39.92}} & \textbf{39.26} & 37.64 & \cellcolor{gray!7}\textbf{38.54}   & \multicolumn{1}{c}{\textbf{2.36}} & \textbf{2.40}  & \multicolumn{1}{c}{47.01} & 47.37   \\ 
\textbf{P1}    & \checkmark& & \multicolumn{1}{c}{40.89} & 33.07 & \cellcolor{gray!7}{36.98} & 36.13 & 29.62 & \cellcolor{gray!7}32.87   & \multicolumn{1}{c}{\textbf{2.36}} & 2.48  & \multicolumn{1}{c}{48.96} & 50.99   \\ 

\textbf{P2} &&  \checkmark & \multicolumn{1}{c}{ \textbf{41.48}} & \textbf{37.88}    & \cellcolor{gray!7}{39.68} &  \multicolumn{1}{c}{31.03} &  \textbf{41.01}  & \cellcolor{gray!7}{36.02} &  2.37  & 2.43 & 54.51 & 53.72    \\ 

\textbf{P3}  & \checkmark& \checkmark  & \multicolumn{1}{c}{35.68} & 29.17  &\cellcolor{gray!7}{32.43} & 30.01 & 31.40  & \cellcolor{gray!7}{30.71}   & \multicolumn{1}{c}{2.41} & 2.51  & \multicolumn{1}{c}{\textbf{54.78}} & \textbf{57.93}    \\ 

\bottomrule 

\end{tabular}
\end{table*}

\begin{figure}
  \centering
\includegraphics[width=5in]{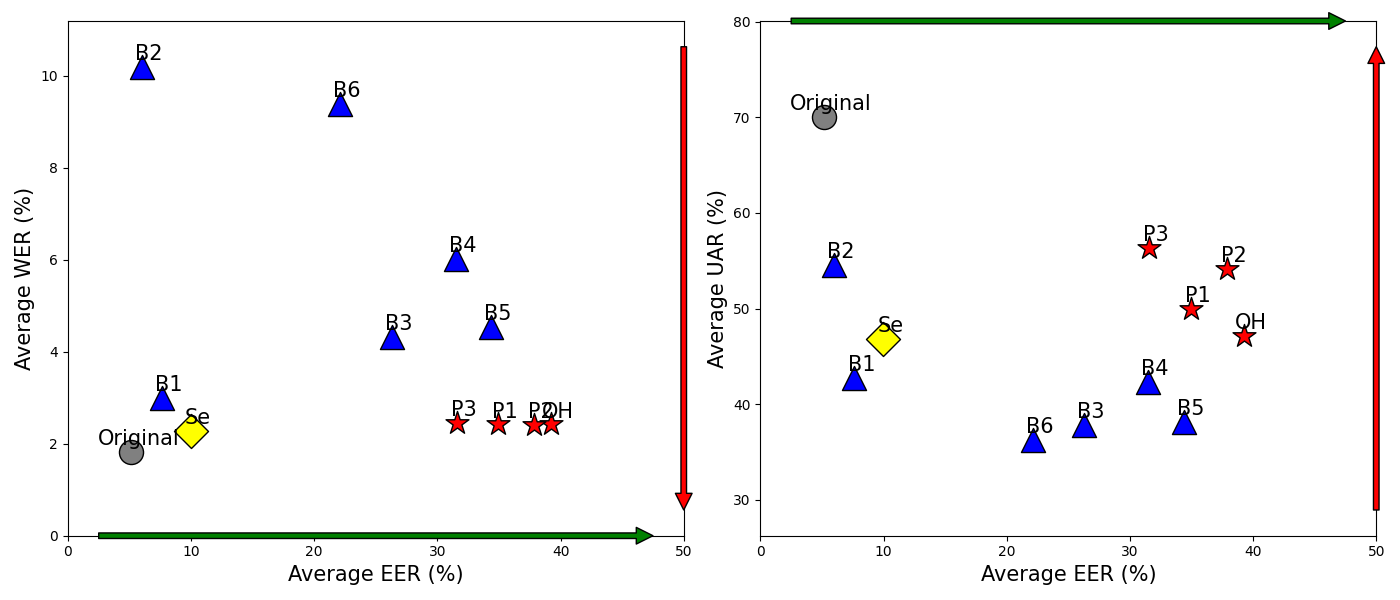}
    \caption{The EER-WER (left) and EER-UAR (right) results for baseline SASs and proposed SASs. Blue triangles correspond to VPC 2024 baseline systems, the grey circles to the original (unprocessed) data, and the red stars to the proposed \textbf{OH} series SASs. The green arrows represent that higher EERs indicate better privacy protection, while the red arrows represent that lower WER and higher UAR indicate better utility.}\label{fig:overall-results}
\end{figure}

\subsubsection{Overall Performance Comparison}
After evaluating the effectiveness of emotion preservation capabilities for the proposed approaches, we move on to the evaluation of overall performance in terms of utility and privacy. Table \ref{tab:propose-compare} lists the EER, WER, and UAR for \textbf{Original}, \textbf{OH}, \textbf{P1-P3}.
From the results, the first observation is that for EER, \textbf{OH} and \textbf{P2} have similar performance and outperform \textbf{P1} and \textbf{P3}. All of them obtain an EER above 30\%, indicating reasonable privacy protection ability.
For WER, all the proposed SASs achieve similar WERs around 2.4\% with no significant differences, approaching the original speech, which achieved around 1.83\%, indicating good speech content preservation.
Another interesting point is that \textbf{P3} achieves the best UAR while having the worst EER. This illustrates that achieving a good trade-off between utility and privacy in the speaker anonymization task is challenging.

Figure~\ref{fig:overall-results} plots the averaged EER-WER (left) and EER-UAR (right) for all the SASs, including both the baselines and proposed systems.
Generally, the proposed systems (red stars) obtain better privacy and utility in terms of both speech content and emotion compared to the VPC baselines. Among the proposed \textbf{OH} series systems, using the solely emotion compensation strategy (\textbf{P2}) is more effective than solely using the emotion encoder (\textbf{P1}), though combining both strategies can further improve emotion preserving performance (\textbf{P3}). \textbf{P2} achieves the best trade-off between EER-UAR and EER-WER.

\begin{table}[h!]
\caption{UAR (\%) obtained by various emotion classifiers trained on different speaker embeddings.}

\centering
\begin{tabular}{lcc}
\hline
\textbf{Speaker embeddings} & \textbf{ First Epoch Model} & \textbf{Final Epoch Model} \\
\hline
Orignal   &62.33  & 69.19   \\
+ OHNN    &  56.27& 65.04  \\
+ OHNN + EC  &90.87 & 90.87   \\
\hline
\end{tabular}
\label{tab:spk-emo}
\end{table}

\begin{figure}
  \centering
\includegraphics[width=3in]{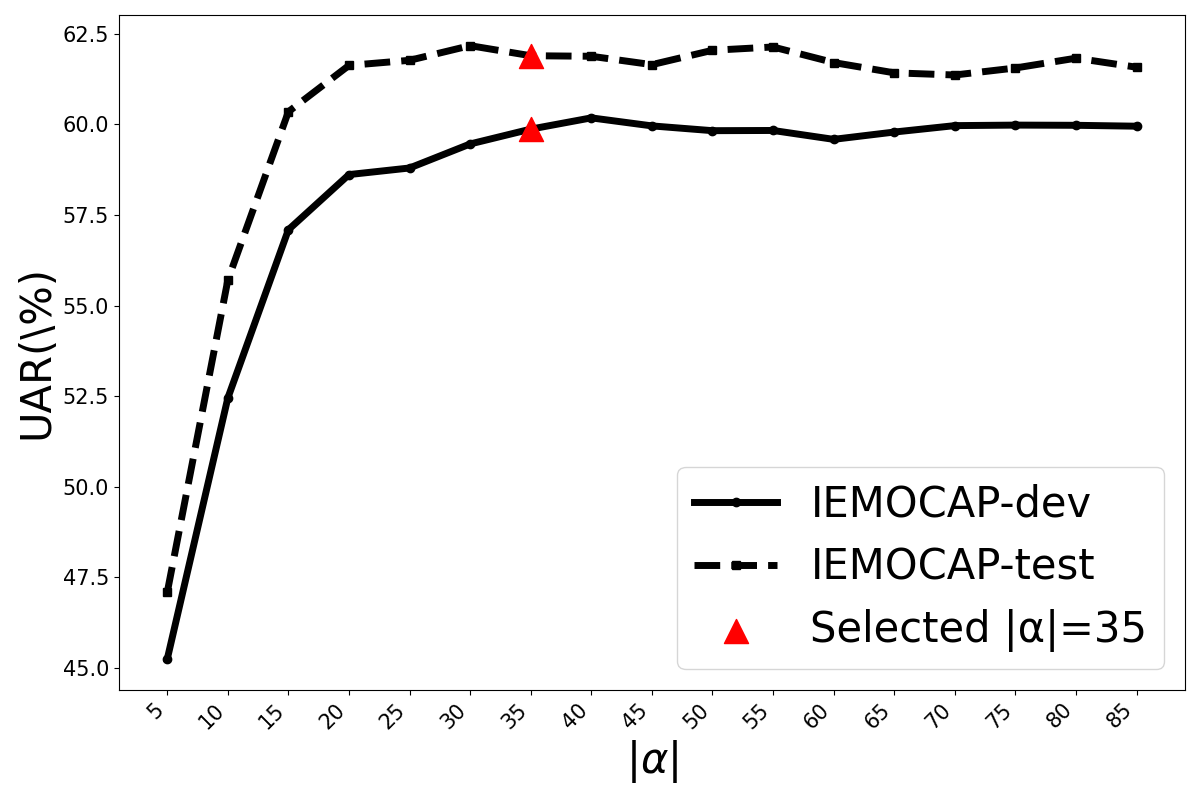}
    \caption{UAR results for anonymized speech processed with emotion compensation manipulation using different $\alpha$.}\label{fig:alpha}
\end{figure}

\begin{table}[h]
\caption{UAR (\%) and recall (\%) for each emotion achieved on SAVEE dataset processed by \textbf{OH} and variants of \textbf{OH} with the emotion encoder (EE) and emotion compensation (EC).}
\label{tab:savee}
\centering
\footnotesize
\begin{tabular}{l  c cccc} 
\toprule
 & {UAR} & {Sad} & {Neutral} & {Angry} & {Happy} \\ 
\midrule
\textbf{Ori.} &\cellcolor{gray!7}{95.33}	&91.67	&99.17	&100.00	&86.67 \\
\midrule
\textbf{OH}  & \cellcolor{gray!7}{63.33} &	50.00	&95.83	&43.33	&31.67 \\
\textbf{P1}  & \cellcolor{gray!7}{63.67} &	50.00	&\textbf{99.17}	&33.33	&36.67\\
\textbf{P2}  & \cellcolor{gray!7}{\textbf{76.67}} &	\textbf{66.67} & 92.50 & \textbf{93.33} &	38.33\\
\textbf{P3}  & \cellcolor{gray!7}{67.33} &	43.33 & \textbf{99.17} &	53.33 & \textbf{41.67} \\
\bottomrule
\end{tabular}
\end{table}
\normalsize

\section{Discussions}
\label{sec:disscussion}

\textit{How much emotional information is retained within speaker embeddings?}
In this work, we assume that original speaker embeddings contain emotional information, as demonstrated in \citep{ulgen2024revealing}. We also conducted an experiment to verify this assumption by using various speaker embeddings to train an emotion classifier and comparing the UAR achieved on the development set.
Specifically, we combined the development and test sets of \textit{IEMOCAP} and split them into 85\% training and 15\% development sets. We trained an emotion model on different speaker embeddings using two fully connected layers: the first maps the 192D speaker embeddings to 128D, and the second maps them to four emotion categories, with a softmax cross-entropy loss function. Early stopping was applied if the development loss did not change for 10 epochs.

We evaluated three types of speaker embeddings, collecting results from both the first epoch model and the final epoch model. The UAR results are presented in Table \ref{tab:spk-emo}, where results using speaker embeddings extracted from the original speech, processed through the OHNN speaker anonymizer, and after further applying emotion compensation are displayed in the first, second, and third lines, respectively.
We see that the original speaker embeddings yield 69.19\% UAR after full training, and 62.33\% at the initial stage, indicating that some emotional clues are present in the speaker embeddings. After applying OHNN, the UAR decreases because OHNN transforms the speaker embeddings based on speaker identity, neglecting emotion preservation. However, with the further application of emotion compensation, the UAR increases to 90.87\% when using both the first and final models.

These results not only confirm that original speaker embeddings contain emotional information but also show trends similar to those observed in Table \ref{tab:ser-results}, where adapting the emotion compensation approach is effective in maintaining the emotional trait.

\textit{How UAR performance changes with different $\alpha$ values in emotion compensation?}
Fig. \ref{fig:alpha} plots the UAR results for anonymized speech processed with emotion compensation manipulation using different $\alpha$. Note that as explained in section \ref{sec:sys-set}, we use the real labels from \textit{IEMOCAP} to determine $\alpha$. Since there are no cascade errors for selecting the correct emotion SVM, the UARs are always higher compared to those using predicted emotion by the emotion indicator.
The results show that varying $\alpha$ does not significantly affect the outcomes for both \textit{IEMOCAP-dev} and \textit{IEMOCAP-test}, indicating the robustness of the emotion compensation strategy.
In other words, the SVMs and adjusted $\alpha$ values can be kept the same across different datasets, which facilitates the practical application of the proposed approach.

\textit{How well is emotion preserved on a completely unseen dataset?} We applied the $\alpha$ values selected on the IEMOCAP dataset directly to the SAVEE dataset~\citep{savee} to generate anonymized speech. Instead of using $SER_\text{eval}$, we used emo2vec+\footnote{\url{https://huggingface.co/emotion2vec/emotion2vec_plus_base}} as an alternative evaluation model, denoted $SER_\text{eval}^\text{emo2vec+}$\footnote{We initially used $SER_\text{eval}$ and observed that the recall for the sad emotion from the original SAVEE speech was 0, indicating a strong mismatch in the $SER_\text{eval}$ model. To address this mismatch caused by the evaluation model's training data, we then used $SER_\text{eval}^\text{emo2vec+}$.}. The results are presented in Table \ref{tab:savee}. We observed a trend where the UAR values follow the order $\textbf{P2} > \textbf{P3} > \textbf{P1} > \textbf{OH}$, indicating that both emotion compensation and the emotion indicator improve emotion preservation on this unseen SAVEE dataset. Moreover, emotion compensation proved to be more effective.

\section{Conclusion and Future Work}
This work explored how to modify a general disentanglement-based speaker anonymization system when a new speech attribute (i.e., emotion) needs to be preserved more effectively.
We first demonstrated that extracting emotion embeddings from a pre-trained emotion encoder is able to preserve emotional cues, though this method does slightly compromise privacy protection.
To further enhance emotion preservation, we introduced an emotion compensation strategy as a post-processing step following the anonymization of original speaker embeddings. This strategy aims to modify anonymized speaker embeddings to reintroduce and preserve emotional traits that may have been lost during the anonymization process. 

Resuts presented here clearly demonstrate that the proposed approach mitigates the loss of emotional attributes caused by the speaker anonymization process, while ensuring that emotional information is preserved more effectively.
We expect these findings will contribute to addressing the preservation of other attributes in speaker anonymization systems, as well as providing insights that can be applied to various other paralinguistic features.

Future work will focus on developing a multiple paralinguistic emotional compensation strategy, aiming to modify one attribute without influencing other attributes.

\clearpage

\bibliographystyle{cas-model2-names}

\bibliography{cas-refs}

\end{document}